\documentclass[prl,twocolumn,showpacs,preprintnumbers]{revtex4-1}

\usepackage{graphicx}
\usepackage{dcolumn}
\usepackage{bm}
\usepackage{amsmath,amssymb}

\begin{document}


\title{Quantum rainbow scattering at tunable velocities}

\author{M. Strebel$^1$}
\author{T.-O. M\"uller$^2$}
\author{B. Ruff$^1$}
\author{F. Stienkemeier$^1$}
\author{M. Mudrich$^1$}
\affiliation{$^1$Physikalisches Institut, Universit\"at Freiburg, 79104 Freiburg, Germany}
\affiliation{$^2$Physik Department T30a, Technische Universit\"at M\"unchen, 85747 Garching, Germany}
\date{\today}

\begin{abstract}
Elastic scattering cross sections are measured for lithium atoms colliding with rare gas atoms and SF$_6$ molecules at tunable relative velocities down to $\sim50\,$m/s. Our scattering apparatus combines a velocity-tunable molecular beam with a magneto-optic trap which provides an ultracold cloud of lithium atoms as a scattering target. Comparison with theory reveals the quantum nature of the collision dynamics in the studied regime, including both rainbows as well as orbiting resonances.
\end{abstract}

\maketitle

The quantum mechanical aspects of atomic and molecular collisions have fascinated physicists as well as chemists since the advent of atomic and molecular beam experiments~\cite{Bernstein:1966}. At low collision energies, rainbow features, glory oscillations and orbiting resonances determine the collision cross sections~\cite{Skodje:2000,Toennies:2007,Chandler:2010}. In the early molecular beam studies, measuring interference and resonance structures allowed to precisely determine interatomic potential parameters. 

While conventional crossed molecular beam experiments have mostly been restricted to high collision energies $\gtrsim 100\,$K$\times k_{\mathrm{B}}$, a variety of cooling and trapping techniques for atoms and molecules have been developed in the last decades~\cite{Krems:2009,Smith:2008}. Using either trapped atoms or trapped molecules or both, atom-molecule collisions can now be explored at collision energies $\lesssim 1\,$K~\cite{Gilijamse:2006,Sawyer:2008,Hummon:2011,Parazzoli:2011, Narevicius:2012}.
While collision studies with trapped species allow for very low collision energies, experimental observables are mostly limited to the integral cross section inferred from trap loss measured in a narrow energy interval.

In this Letter we present measurements of the cross section for elastic collisions of lithium (Li) atoms with rare gas (Rg) atoms and sulfur hexafluoride (SF$_6$) molecules over a wide range of collision energies $2\lesssim E_{\text{coll}}\lesssim 200\,$K$\times k_{\mathrm{B}}$. Our setup comprises a magneto-optical trap (MOT) for Li atoms and a molecular beam source which produces slow beams of atoms and molecules by jet expansion out of a fast rotating nozzle~\cite{Gupta:1999,Strebel:2010}. This approach combines the advantages of a MOT for producing a dense ultracold atomic scattering target with the benefits of the rotating nozzle technique of providing velocity-tunable cold beams of atom or various chemically relevant molecules. Besides, scattered Li atoms or Li containing product compounds can be sensitively detected by simple surface ionization. Since Li is the lightest of all alkali atoms it is particularly well suited for quantum scattering experiments. Its chemical reactivity will allow low-energy reactive scattering studies in the future. By detecting the trap loss out of the MOT as well as scattering products in the direction of the molecular beam, both integral as well as backward differential scattering cross sections can be derived. In this way, multiple rainbow scattering features in the energy dependence of the cross section for Li-Rg and Li-SF$_6$ backward scattering can be resolved at previously inaccessible low energies. Full quantum scattering calculations reveal a rich structure of rainbow and orbiting resonances in the relevant energy range.

\begin{figure}
\begin{centering}
\includegraphics[width=7.5cm]{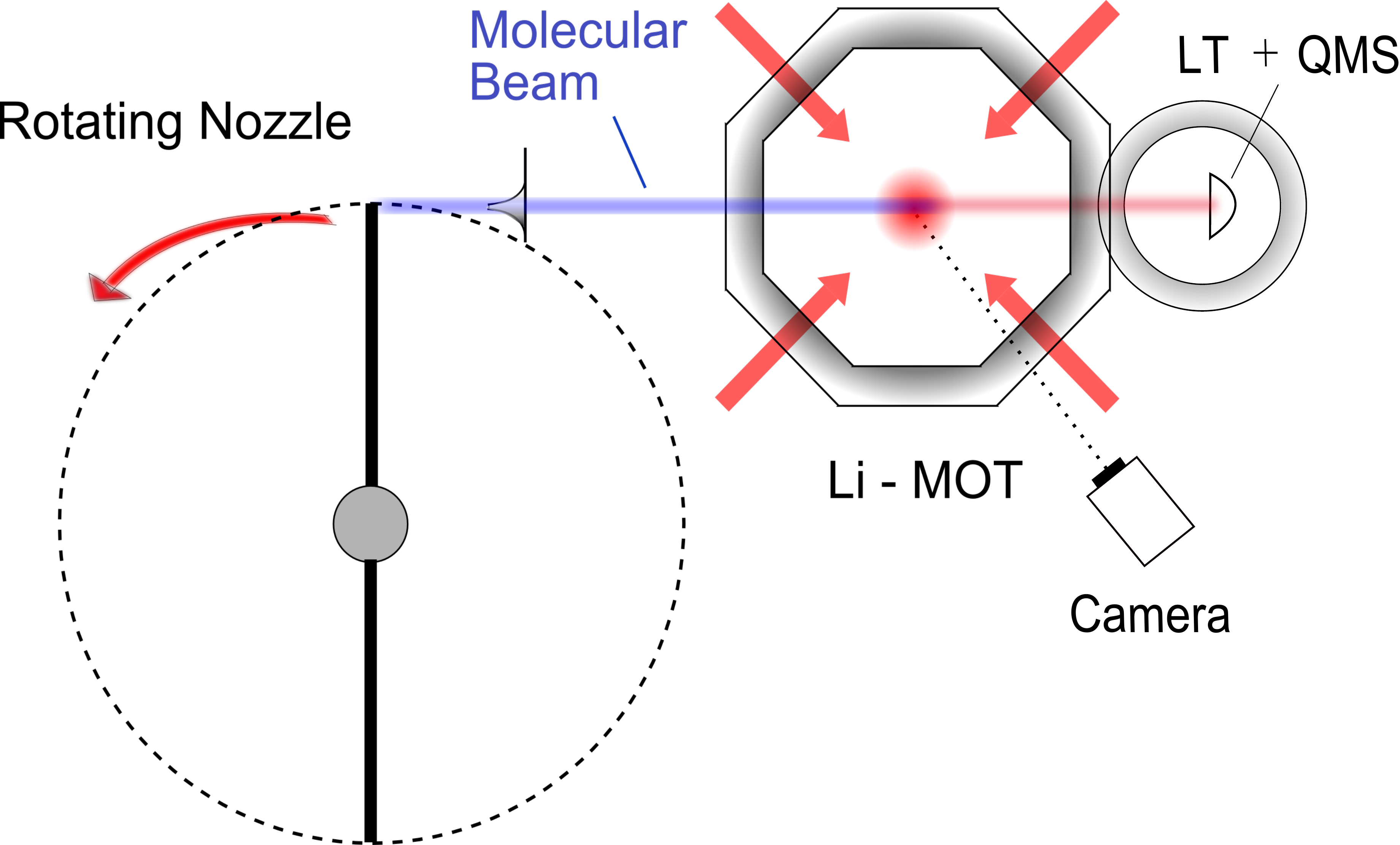}
\end{centering}
\caption{(Color online) Sketch of the experimental setup consisting of a nozzle mounted at the tip of a fast spinning rotor (left part) and a magneto-optic trap (MOT) for lithium atoms (right part). A quadrupole mass spectrometer (QMS) and a Langmuir-Taylor (LT) detector are placed behind the MOT in the molecular beam direction for detecting the primary nozzle beam and the scattered atoms, respectively. \label{fig:Setup2} }
\end{figure}
The scattering apparatus is schematically represented in Fig.~\ref{fig:Setup2}. The source setup for producing velocity-tunable supersonic jets is sketched on the left hand side. Since it has been described in detail we only briefly summarize the key features here~\cite{Strebel:2010}. The principle of operation relies on the mechanical translation of the nozzle orifice at high speeds $v_{\text{rot}}$ in either forward or backward direction. The molecular beam that exits the nozzle at a velocity $v_s$ is then accelerated or decelerated, respectively, to the speed $v_0=v_\text{s}\pm v_\text{rot}$ in the laboratory (LAB) frame~\cite{Gupta:1999}. By spinning the rotor with radius $R=19\,$cm at frequencies up to 350\,Hz (21000\,rpm) we obtain pulsed beams in the speed range $50\lesssim v_0 \lesssim 1000\,$m/s with a longitudinal velocity spread corresponding to a beam temperature in the range $2\lesssim T_\text{beam} \lesssim 50\,$K. One main advantage of the technique besides its conceptual simplicity is its universal applicability to any type of atomic or molecular gas that has sufficient vapor pressure at room temperature~\cite{Strebel:2010,Strebel:2011}.

This molecular beam line is connected to a compact magneto-optical trap setup for Li atoms which is extended by an additional detector chamber, sketched on the right hand side of Fig.~\ref{fig:Setup2}.
The laser system is diode laser-based and a decreasing field Zeeman slower is used for loading the MOT at a rate of about $10^8$ atoms per second. At the laser detuning from the $2s_{1/2}\rightarrow 2p_{3/2}$ atomic transition frequency of $\delta=8.5\gamma$, where $\gamma\approx 5.9$\,MHz denotes the natural line width, we typically load up to $10^9$ Li atoms into the MOT at a temperature $T_\text{MOT}\sim1\,$mK. We estimate the fraction of excited atoms in the MOT to $s_0/2(1+s_0+(2\delta/\gamma)^2)\approx 11\%$, where $s_0\approx 35$ denotes the saturation parameter given by the laser intensity~\cite{Metcalf:1999}.

\begin{figure}
\includegraphics[width=7.5cm]{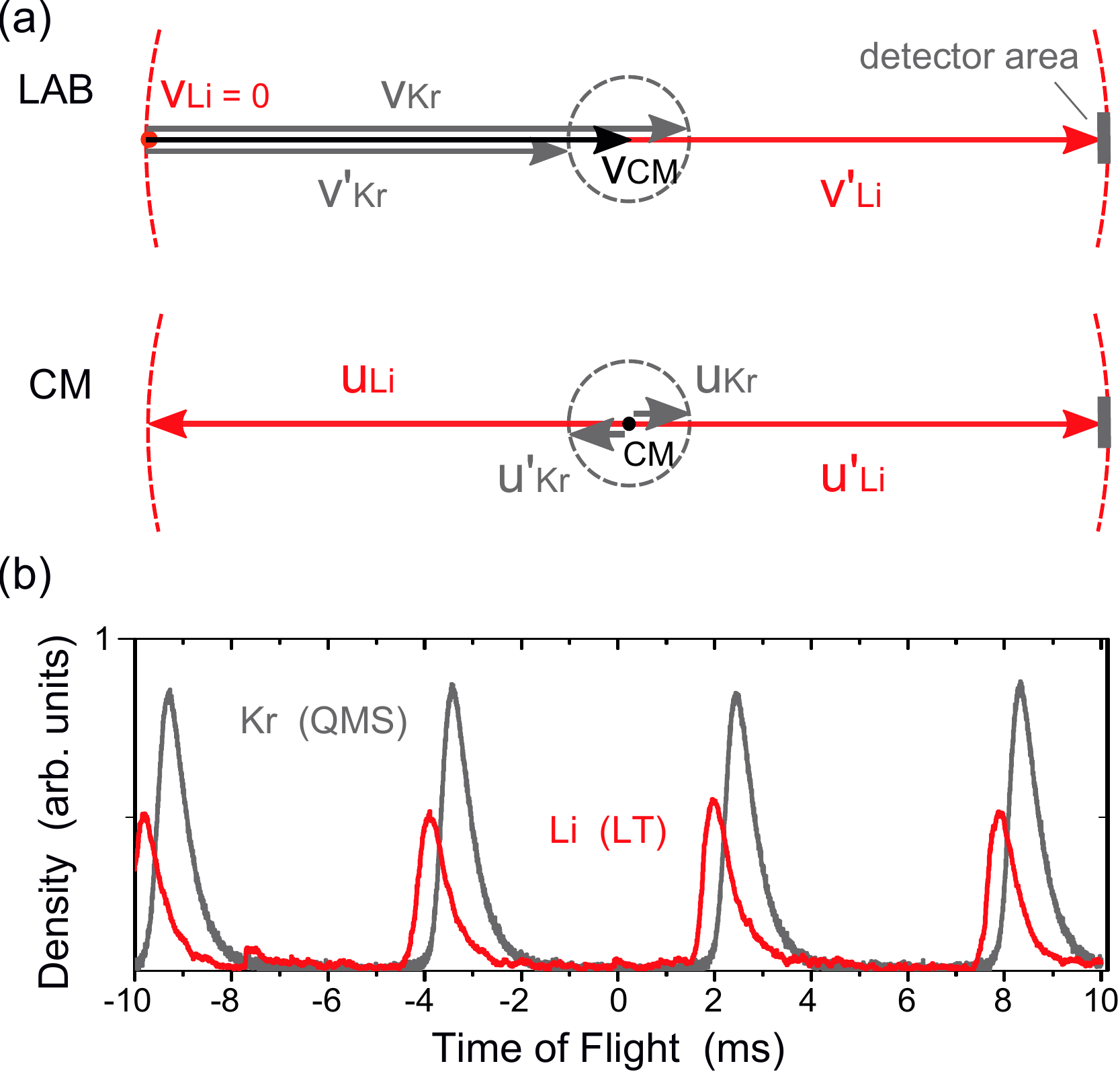}
\caption{\label{fig:Timeofflight} (Color online) (a) Newton diagrams in the laboratory (LAB) and center of mass (CM) frames for the case of head-on collisions of slow krypton (Kr) atoms in the nozzle beam and resting Li atoms in the trap. (b) Time of flight distributions of slow Kr beam pulses out of the rotating nozzle and elastically scattered Li atoms. }
\end{figure}

Three different detectors are used for performing scattering measurements. A CCD camera monitors the position and total fluorescence of the MOT atoms from which we infer the total number of trapped Li atoms $N_\text{MOT}$. A commercial quadrupole mass spectrometer (QMS) and a Langmuir-Taylor (LT) detector are mounted behind the MOT on the axis of the nozzle beam. The QMS is used for characterizing the primary beam in terms of density $n_\text{beam}$ and velocity $v_0$. Fig.~\ref{fig:Timeofflight} (b) shows a typical time of flight density trace of a krypton (Kr) beam with velocity $v_0=190\,$m/s recorded with the QMS (gray line). The LT detector is based on surface ionization of alkali atoms when hitting the surface of a hot Rhenium (Re) ribbon~\cite{Delhuille:2002}. The red line in Fig.~\ref{fig:Timeofflight} (b) depicts the transient LT detector signal $S$ which is proportional to the flux of scattered Li atoms $\dot{N}_\text{Li}$ out of the MOT due to elastic head-on collisions, i.\,e. $S=k\dot{N}_\text{Li}$. The detection efficiency ($k \approx 0.01$) is inferred from the comparison with theoretical calculations. Thus, the differential cross section in beam direction is determined by
\begin{equation}
\label{eq:sigmaDifferential}
\frac{d\sigma}{d\Omega} \left(v_0,\theta=180^\circ \right)=\frac{\dot{N}_{\text{Li}}}{v_{\text{0}}\,n_{\text{beam}}\,N_{\text{MOT}}}\,.
\end{equation}

The collision kinematics is illustrated in the Newton diagrams in Fig.~\ref{fig:Timeofflight} (a) which relate velocity vectors in the LAB and center of mass (CM) frames. In the LAB frame the Li atoms are initially at rest compared to the Kr atoms, $\vec{v}_\text{Li}\approx\vec{0}$, and the CM velocity $\vec{v}_\text{CM}$ is mostly given by the velocity of the heavier Kr atom, $\vec{v}_\text{CM}\approx\vec{v}_\text{Kr}$. In the CM system, this translates to a Li velocity $\vec{u}_\text{Li}$ which is reversed during the collision, $\vec{u'}_\text{Li}=-\vec{u}_\text{Li}$. Thus, the LT detector measures backward scattered Li atoms, a situation more difficult to achieve in standard crossed beam experiments. The velocity of scattered Li in the LAB system, $\vec{v'}_\text{Li}$, points into the Kr beam direction with a magnitude of roughly twice the initial Kr beam velocity, $\vec{v'}_\text{Li}\approx2\vec{v}_\text{Kr}$, which explains the time slip between the two traces in Fig.~\ref{fig:Timeofflight} (b).

During the measurements the nozzle pressure is actively stabilized to $p_{0}=0.15\,$bar taking into account the pressure rise by centrifugal forces at high speeds of rotation~\cite{Gupta:1999,Strebel:2010}. The MOT is operated continuously which implies that the loading rate $L$ equals the total trap loss rate $R=R_{\text{beam}}+R_{\text{bg}}$. The latter results from losses due to nozzle beam scattering $R_{\text{beam}}$ and residual gas collisions $R_{\text{bg}}$. Since both the nozzle beam intensity and the background pressure in the MOT chamber $p_{\text{bg}}$ sensitively depend on the direction and speed of rotation of the nozzle, we infer $L$ and $R_{\text{bg}}(p_\text{bg})$ from separate measurements of MOT loading curves as a function of $p_\text{bg}$. The integral cross sections $\sigma_{\text{tot}}(v_0)$ are then given by
\begin{equation}
\label{eq:sigmaIntegrated}
\sigma_{\text{tot}}(v_0)=\frac{R_{\text{beam}}(N_{\text{MOT}},p_\text{bg})}{v_0\,n_{\text{beam}}\,N_{\text{MOT}}},
\end{equation}
where $R_{\text{beam}}(N_{\text{MOT}},p_\text{bg})=L/N_{\text{MOT}}-R_\text{bg}(p_\text{bg})$.

\begin{figure}
\includegraphics[width=8cm]{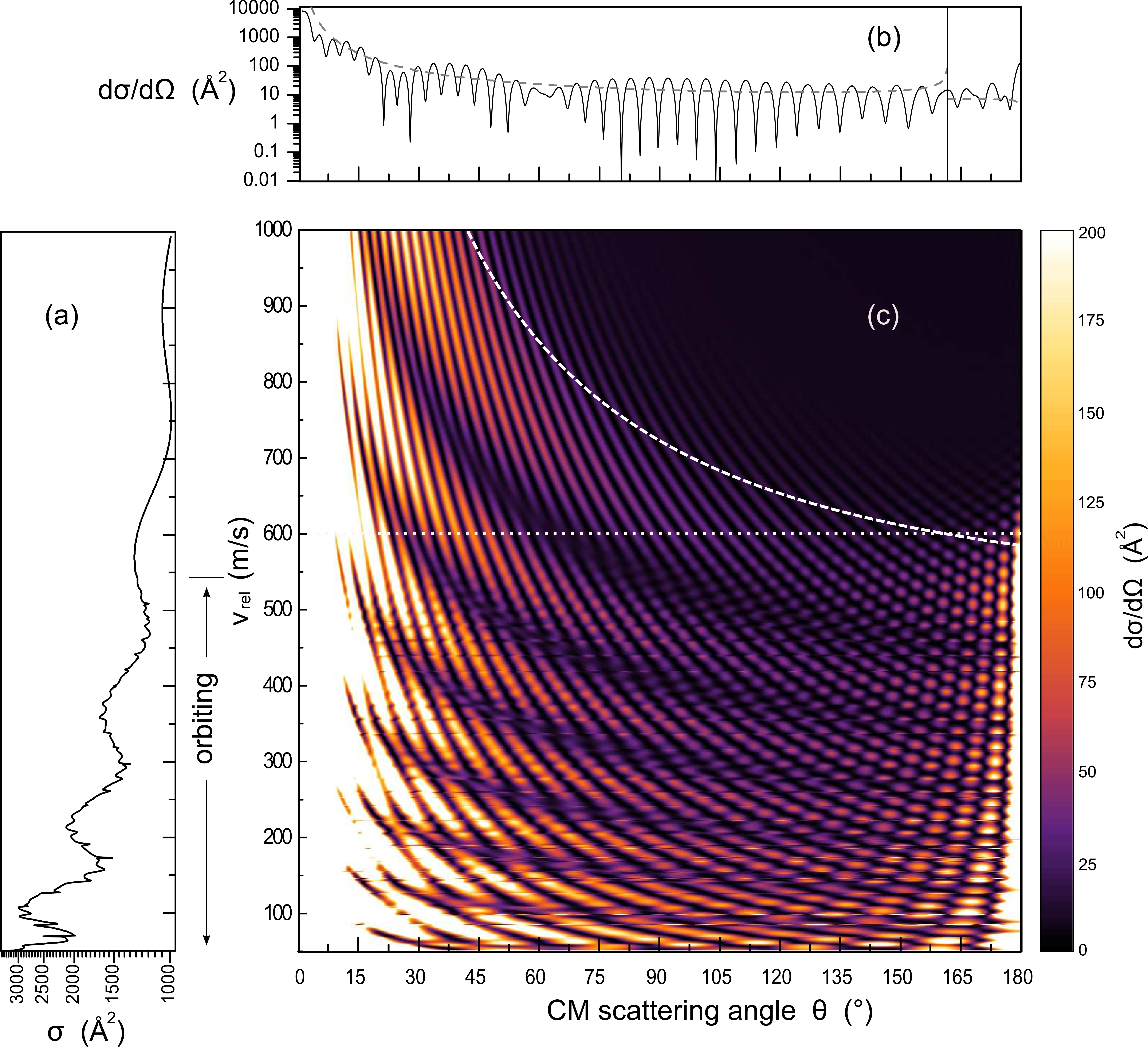}
\caption{\label{fig:2Dplot} (Color online) Calculated cross sections for elastic scattering of Li and Xe atoms. (c) Differential cross sections as a function of CM scattering angle and velocity. (b) Angle-differential cross section at fixed velocity, corresponding to a cut through (c) along the dotted line. The dashed line depicts classical calculations. (a) Integral cross section obtained by integrating over scattering angles. }
\end{figure}
The experimental Li-Rg elastic scattering cross sections are compared with full quantum scattering calculations based on Lennard-Jones interaction potentials $V(r)=\epsilon\, ((r_{\text{m}}/r)^{12}-2(r_{\text{m}}/r)^6)$. Here, $\epsilon$ and $r_{\text{m}}$ denote the depth and position of the potential well, respectively. 
The scattering phase shifts are calculated numerically by solving the Schr\"odinger equation for partial waves $l=0$ up to $l=200$ using Numerov's algorithm.
The resulting differential cross section for ground state Li-xenon (Xe) scattering is displayed in Fig.~\ref{fig:2Dplot}~(c) as a function of scattering angle $\theta$ and relative velocity $v_0$. The intensity pattern is determined by interfering slow and fast intensity modulations and by pronounced maxima in forward $\theta=0^\circ$ and backward $\theta=180^\circ$ directions at low $v_0$. Fig.~\ref{fig:2Dplot}~(b) shows the angle-differential cross section $d\sigma/d\Omega (\theta)$ at fixed relative velocity obtained by a cut through the data in (c) at $v_0=600$\,m/s as indicated by the dotted line. The broad maximum at large scattering angles around $\theta=100^\circ$ can be associated with the main rainbow that is reminiscent of the classical rainbow, i.\,e., the divergence of $d\sigma/d\Omega$ at $\theta=162^\circ$ as obtained from the corresponding classical calculation, shown as dashed line in Fig.~\ref{fig:2Dplot}~(b). Its low-energy replicas are referred to as supernumerary rainbows~\cite{Bernstein:1966}. They are pure quantum phenomena that have no classical analog. The fast (or secondary) interference oscillations are related to distant classical trajectories of the projectile around the target atom that still end up at the same solid angle. The large difference between the quantum and classical calculations highlights the quantum nature of the scattering dynamics even at energies $E_\text{coll}\sim\epsilon=140\,$K$\times k_\mathrm{B}$ ($v_0=585$~m/s). The systematic shifting of the rainbow features in Fig.~\ref{fig:2Dplot}~(c) toward $\theta=180^\circ$ as $E_\text{coll}$ is reduced to $\epsilon$ closely follows the classical rainbow divergence shown as thick dashed line. In analogy to classical glory scattering, the backward scattering intensity rises sharply as $E_\text{coll}$ falls below $\epsilon$. This rise can be seen as the addition of scattering intensity from the attractive part of the interaction potential to the one from the repulsive hard core potential. For $E_\text{coll}\gg\epsilon$ the backward scattering cross section approaches the classical value $R^2/4$ where $R$ denotes the radius of the inner repulsive core.

The integral cross section, shown in Fig.~\ref{fig:2Dplot}~(a), is obtained by integrating $d\sigma/d\Omega$ over the full solid angle for each value of $v_0$. The slow modulations (glory oscillations) are predominantly caused by the rainbow maxima at small angles $\theta\lesssim10^\circ$ (bright areas in Fig.~\ref{fig:2Dplot}~(c)). At even lower scattering energies $E_\text{coll}\lesssim0.8\epsilon$ ($v_0\lesssim 523\,$m/s), additional narrow quantum mechanical resonance features appear in Fig.~\ref{fig:2Dplot}~(a) and (c). These stem from predissociative rotational states of the scattering complex with resonantly prolonged lifetimes, that may be viewed as the quantum version of classical orbiting~\cite{Chandler:2010}.

\begin{figure}
\includegraphics[width=8cm]{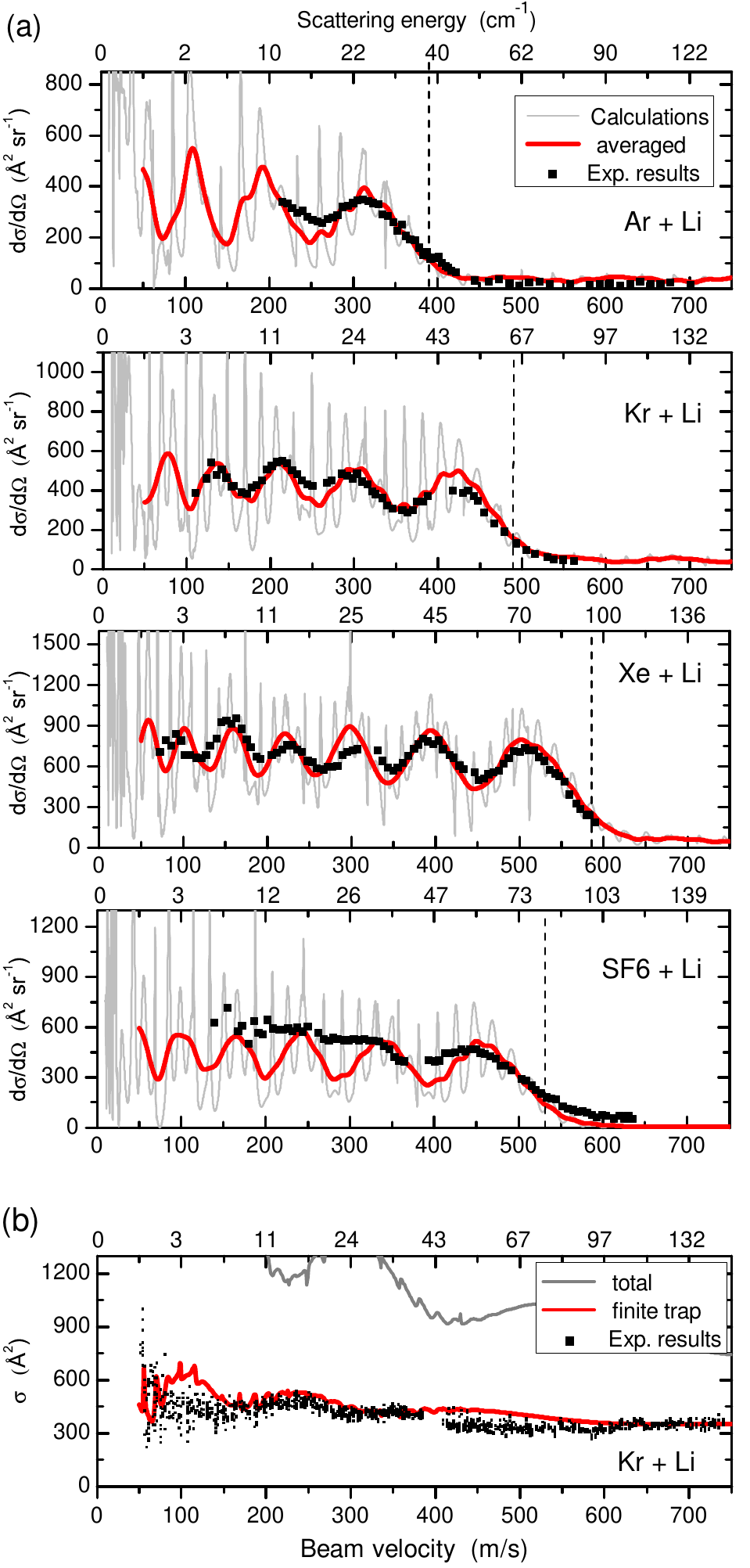}
\caption{\label{fig:differential} (Color online) (a) Experimental (symbols) and calculated (lines) differential cross sections of Ar, Kr, Xe, and SF$_6$ elastic backward scattering with trapped Li atoms as a function of relative velocity. (b) Integral cross sections for Li+Kr collisions.}
\end{figure}

Fig.~\ref{fig:differential} (a) shows typical elastic differential scattering cross sections for backward scattering of argon (Ar), Kr, Xe, and SF$_6$ with Li as a function of beam velocity $v_0$ as inferred from the LT-detector signals according to Eq.~(\ref{eq:sigmaDifferential}) (symbols). The corresponding CM collision energy is indicated on the top scales of each panel.
In all measurements we observe a clear onset for backward scattering intensity at $v_{0,\text{on}}\approx$387, 490, 585, and 530\,m/s, respectively, followed by an oscillatory structure at lower $v_0$. This intensity onset coincides with the appearance of the classical rainbow divergence in backward scattering from ground state Li at $v_{0,\text{rb}}$ (vertical lines). Thus, we identify the maximum at highest $v_0$ as the main rainbow in backward direction and the additional features at lower $v_0$ as backward supernumerary rainbows measured as a function of collision energy. The increase of the onset velocity $v_{0,\text{on}}$ with rising Rg atom number results from the direct correlation of $v_{0,\text{on}}$ to the potential well depth $\epsilon$. The latter is mostly determined by the dynamic polarizability of the Rg atom which increases from Ar to Xe. Measurements using CF$_3$H and CO$_2$ molecules show less pronounced intensity onsets at $v_{0,\text{on}}\approx$550 and 650\,m/s, respectively.

The results of the quantum scattering calculations are shown as thin solid lines. Elastic collisions from excited Li atoms, which are taken into account except for the Li+SF$_6$ case, make a notable contribution only at velocities $v_0\gtrsim v_{0,\text{rb}}$.
The thick solid line is obtained when allowing for the experimental energy resolution which is mainly limited by the velocity spread of the Rg beam, $\Delta v\approx 40$\,m/s. The agreement with the experimental data is satisfactory with slight deviations around those beam velocities attained at nozzle speeds $v_{\text{rot}}\approx 0$ ($v_0=400$\,m/s for Li+Kr, $v_0=320$\,m/s for Li+Xe). At these values of $v_0$, the determination of beam density $n_{\text{beam}}$ is inaccurate due to distortions of the beam pulses by the skimmer opening function~\cite{Strebel:2010}. Note that in the case of Li+SF$_6$, reactive collisions may add to the scattering signal~\cite{parrish:2518}.

Fig.~\ref{fig:differential}~(b) displays measured integral scattering cross sections $\sigma(v_0)$ for Li+Kr collisions obtained from the MOT fluorescence using Eq.~(\ref{eq:sigmaIntegrated}). We observe nearly constant $\sigma\approx 320\,$\AA$^2$ over the full range of velocities, with a slight increase and weak modulations as $v_0$ is tuned below 500\,m/s. The calculated integral scattering cross sections, shown as a gray line, significantly exceed the measured values in both magnitude and contrast of modulations due to glory oscillations. This discrepancy is due to the perturbing effect of the Li MOT on collision-induced losses, as detailed in~\cite{Fagnan:2009}. When taking into account the estimated effective depth of the MOT trapping potential $U_\text{trap}\approx 0.62\,$K$\times k_\mathrm{B}$, we obtain the result shown as a red line, which matches the experimental values much better. Note that the differential cross sections (Fig.~\ref{fig:differential}~(a)) are not affected, since the transferred energy in head-on collisions by far exceeds $U_\text{trap}$ in the studied range of $v_0$.

In summary, we have presented a scattering apparatus that allows us to measure in a wide range of relative velocities both integral as well as differential cross sections for collisions of atoms and molecules in a nozzle beam with trapped ultracold Li atoms. We find that measuring differential scattering signals, in particular in backward direction using a scattering detector, is much better suited for resolving the details of the collisions dynamics than merely detecting trap loss. We thus clearly evidence elastic multiple quantum rainbow structures in backward direction down to collision energies $E_{\text{coll}}\approx 2$\,K$\times k_{\mathrm{B}}$. Our approach opens the way to a variety of scattering experiments in a range of collision energies hitherto inaccessible except using cryogenic beam sources of He and H$_2$. The selective detection of elastically scattered Li atoms and Li containing reaction products will allow us to study low-energy reactive collisions in the future.

We gratefully acknowledge support by the Baden-W\"urttemberg-Stiftung and by DFG.


%

\end{document}